\documentclass[aps,prl,twocolumn,final,letterpaper]{revtex4}

\usepackage{graphicx}
\usepackage{import}
\usepackage{epstopdf}
\usepackage{amsmath} 
\usepackage{bm}
\usepackage{amssymb}
\usepackage{quotes}
\usepackage{transparent}
\usepackage{dcolumn}
\usepackage[usenames, dvipsnames]{color}
\usepackage{multirow}
\usepackage{cancel} 
\usepackage{mdframed}
\usepackage{color}
\usepackage{bm}
\usepackage{setspace}
\usepackage{dsfont}
\usepackage{braket}
\usepackage{slashed}
\usepackage{float}
\usepackage{soul, color} 
\soulregister\ref{7}  
\soulregister\cite{7} 
\renewcommand{\st}[1]{}

\newcommand{\rv}{\mathbf{r}}

\newcommand{\kv}{\mathbf{k}}

\newcommand{\gv}{\mathbf{g}}

\newcommand{\appropto}{\mathrel{\vcenter{
  \offinterlineskip\halign{\hfil$##$\cr
    \propto\cr\noalign{\kern.2pt}\sim\cr\noalign{\kern-2.5pt}}}}}

\newcommand{\Av}{\mathbf{A}}
\newcommand{\Ev}{\mathbf{E}}

\newcommand{\Gv}{\mathbf{G}}

\newcounter{myfootnotectr}
\newcommand{\myfootnote}[1]{%
  \stepcounter{myfootnotectr}%
  \textsuperscript{\themyfootnotectr}%
  \footnotetext[\value{myfootnotectr}]{#1}%
}

\begin{document}

\title{Controlling X-ray emission with optical nanostructures}

\author{Elina Sendonaris$^{1,*}$, Jamison Sloan$^{2}$, Nicholas Rivera$^{3,4}$, Ido Kaminer$^{5}$, Marin Solja\v{c}i\'{c}$^{6}$}

\affiliation{$^{1}$Department of Applied Physics, California Institute of Technology, Pasadena, CA 91125, USA \\
$^{2}$Department of Electrical Engineering and Computer Science, Stanford University, Stanford, CA 94305, USA \\
$^{3}$Department of Physics, Harvard University, Cambridge, MA 02138, USA\\
$^{4}$School of Applied and Engineering Physics, Cornell University, Ithaca NY 14853, USA\\
$^{5}$Department of Electrical Engineering, Technion Israel Institute of Technology, Haifa 32000, Israel\\
$^{6}$Department of Electrical Engineering and Computer Science, Massachusetts Institute of Technology, Cambridge, MA 02139, USA\\
$^{*}$esendona@caltech.edu}

\noindent	

\begin{abstract}
Nonlinear processes lie at the heart of many technologies such as frequency converters and entangled photon sources. Historically, observation and manipulation of these processes, for instance through nanostructures, has been limited to optical and lower frequencies. Recently, however, second-order nonlinear processes which couple X-ray and optical photons have been observed and used to probe the electronic structure and optical response of materials. Observing and controlling these processes remains challenging due to their low efficiency and the difficulty of fabricating devices with spatial features on the scale of X-ray wavelengths. Here, we show how optical nanostructures can be used to manipulate X-ray/optical nonlinear processes, using a quantum theory which describes these second-order nonlinear interactions. As an example, we show how photonic crystals shape both the spectral and spatial characteristics of X-rays emitted through X-ray to optical parametric down-conversion, leading to a fill-factor-normalized rate enhancement of 2.2 over an unstructured medium, in addition to control over the directionality of X-ray emission. The ability to control X-ray nonlinear processes may lead to more monochromatic, heralded X-ray sources, enhanced ghost imaging of lattice and electronic dynamics, and imaging and spectroscopy beyond the standard quantum limit. Our framework illuminates a path towards controlling quantum optical effects at X-ray frequencies.
\end{abstract}

\maketitle


Nonlinear optical interactions form the basis for devices which convert between different frequencies of light \cite{si-conv-nonl, vis-telecom-conv}, and they are critical for the design of quantum light sources featuring properties such as entanglement \cite{ent-phot-pairs-spdc, ring-resonator, metasurf-q-phot} and squeezing \cite{vahlbruch2016detection}. However, these nonlinear and quantum processes have historically been limited to microwave-UV frequencies, failing to enter the X-ray regime. This constraint is representative of the broader challenge of controlling X-rays, which generally interact with matter inefficiently. It is highly desirable to extend nonlinear and quantum optics into the X-ray regime in order to better probe atomic physics and utilize low-noise quantum measurement techniques in the X-ray range \cite{xray-q-opt}. The extension of nonlinear optics to the X-ray regime has faced many challenges due to the lack of strong, coherent, monochromatic X-ray sources, low efficiency of X-ray nonlinear processes, and general lack of X-ray nanostructures and devices. 

Despite these challenges, some recent advances have been made in realizing X-ray nonlinear optics. Due to developments in X-ray sources such as synchrotrons, X-ray free-electron lasers (FELs), and high-harmonic generation, experiments have recently demonstrated the first ingredients of X-ray quantum optics. For instance, X-ray parametric down-conversion, a process in which photons interact with vacuum fluctuations to produce entangled photon pairs, has been used to generate X-ray photon pairs \cite{xray-phot-pairs} and perform ghost imaging \cite{xray-ghost-imag}, both with almost no background noise.

\begin{figure}
\centering
\includegraphics[width=\linewidth]{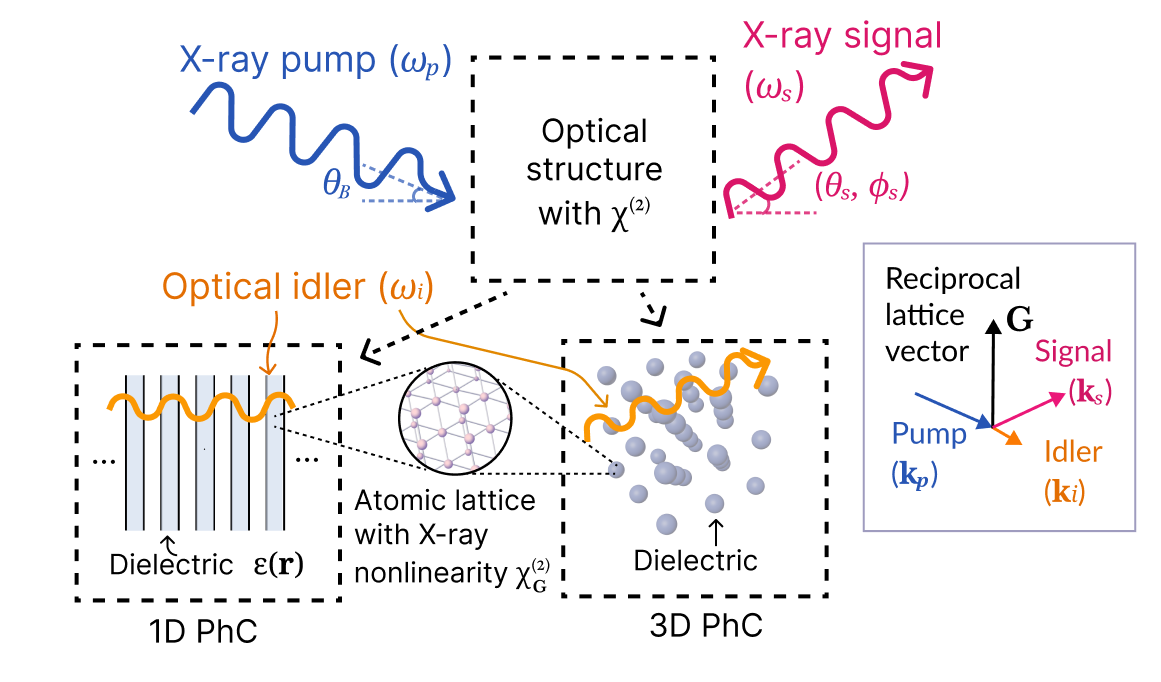}
\caption{X-ray to optical parametric down-conversion (XOPDC) in an arbitrary structure (e.g., a photonic crystal, metamaterial, or other nanostructure). The pump (X-ray) enters at the Bragg angle $\theta_B$ and interacts with an optical structure with an XO nonlinearity $\chi^{(2)}_\Gv$ made of a dielectric with permittivity $\varepsilon$, emitting an optical idler in a mode of the system and an X-ray signal pair at angle $(\theta_s, \phi_s)$. Inset: Diagram of the relevant momentum space vectors involved in XOPDC. Since the idler characteristic wavevector $\kv_i$ is much smaller than the other vectors involved, the momentum conservation condition ($\Gv + \kv_p = \kv_s + \kv_i$) amounts to Bragg diffraction off the reciprocal lattice vector $\Gv$ with a small correction for the idler.}
\label{fig:setup}
\end{figure}

In addition to these processes which couple X-rays to other high frequency photons, there has been success in realizing second order nonlinearities which couple X-ray and optical frequencies. X-ray and optical wave mixing through sum- and difference-frequency generation (SFG and DFG) was proposed over fifty years ago \cite{opt-modulated, mixing-xray-optical, valence-probing}, but was not observed until 2012 \cite{glover}. Numerous other experiments have demonstrated second-order nonlinearities coupling optical/UV and X-ray frequencies and their uses for enhanced imaging and detection \cite{schori, sofer, opt-resp-euv-hres, explain-susc-band-edges}. A particularly interesting process is highly non-degenerate X-ray to optical parametric down-conversion (XOPDC). In this process, an X-ray photon interacts with optical vacuum fluctuations of a crystal to produce an entangled pair: one optical idler photon, and one slightly red-shifted X-ray signal photon (Fig. \ref{fig:setup}). These X-ray-optical (XO) nonlinear processes depend on the density of states of distant frequency ranges, potentially enabling measurement processes which simultaneously combine atomic-scale spatial resolution with optical frequency resolution and detection capabilities. Such processes could eventually enable the probing of atomic and valence electron structures \cite{valence-probing}, as well as the generation of optically heralded X-ray photons.

Although X-ray nonlinear optics has shown some promise, these processes still occur with low efficiencies and offer few degrees of freedom for control. This is in stark contrast to traditional nonlinear optics, which has benefited greatly from developments in materials and photonic nanostructures such as metasurfaces and metamaterials \cite{enh-nonl-phcs, nonlinear-metasurf, metasurf-q-phot, roadmap-metamaterials, nonl-si-phot, zheng2023advances}, for instance via the Purcell effect \cite{tailor-nonl-purcell, mod-spont-emis-nanostruc}. So far, these developments have raised few implications for X-ray processes, since optical materials exhibit very little response at these frequencies. However, the coupling of optical and X-ray densities of states (DOS) through XO nonlinear processes presents an untapped opportunity to influence the behavior of X-rays through the wealth of advances in optical nanostructures.

Here, we show how optical nanostructures can control X-rays through XO nonlinear processes. The rates, frequencies, and directions of emitted X-rays depend on the optical modes and optical DOS due to coupling between optical and X-ray frequencies enabled by XO nonlinearities; thus, optical nanostructures fabricated using XO nonlinear materials directly influence X-ray emission. As an example, we describe how structuring the nonlinear material into photonic crystals (PhCs) can control and enhance XOPDC. In our examples, PhCs enhance XOPDC by 2.2 times at certain frequencies, while suppressing it almost entirely at others, raising the possibility of a strongly spectrally tailored X-ray source. Furthermore, the angular distribution of X-ray emission is linked to the PhC structure. Our work is based on a fully quantum Hamiltonian framework which describes second order XO processes in general optical nanostructures. While we focus on XOPDC in PhCs, this formalism is equally valid for other second-order XO processes (such as SFG and DFG) and optical structures such as metasurfaces and resonant cavities \cite{zheng2023advances}. Our findings illuminate a path which extends quantum optics into the X-ray regime by integrating existing optical nanostructures into new X-ray devices.

The general scheme for controlling XOPDC with optical nanostructures is depicted in Fig. \ref{fig:setup}. In this system, a strong X-ray pump field of frequency $\omega_p$ impinges on an optical nanostructure. The geometry of the nanostructure is encoded through the dielectric function $\varepsilon(\rv)$, from which we can compute the optical modes present in the structure. We show examples of 1D and 3D PhCs, but the concept is generalizable to any nanostructure. Microscopically, an optical field
polarizes the bonds of the crystal lattice, inducing a time-dependent change in the valence charge density $\delta\rho(\rv, t)$ \cite{opt-modulated}. In the absence of this optical field, a pump photon incident at an appropriate angle can elastically Bragg scatter off of the periodic atomic lattice. However, when the charge density is dynamic rather than stationary, the X-ray pump can downconvert into an optical idler photon at frequency $\omega_i$ and an X-ray signal photon at frequency $\omega_s = \omega_p - \omega_i$. This dynamical modulation can either be achieved by a real optical field at frequency $\omega_i$ (as from a laser) or the process can be spontaneous, in which the modulation of the valence charge density occurs due to vacuum fluctuations. Since the vacuum fluctuation density of states at optical frequencies is directly controlled by the optical nanostructure, so is the spectrum of XOPDC emission.

We compute the rate of XOPDC using a Hamiltonian framework which describes second order XO nonlinearities in general photonic nanostructures. Based on the physical interpretation of XO processes described above, we find that the interaction picture Hamiltonian which describes the interaction between the modulated valence charge density and the X-ray vector potential operator $\Av(\rv, t)$ is $H_{\text{int}}(t) = q/(2m) \int_V d^3r~\delta \rho(\rv, t)\mathbf{A}^2(\rv, t)$, where and $q$ and $m$ are the electron charge and mass, and $V$ is the nonlinear region. To explicitly identify the interaction between the three photons in XOPDC (pump, signal, idler), we relate the modulated charge density $\delta\rho$ to the optical idler electric field $\Ev_\text{I}(\rv, t)$ through the polarizability per unit volume $\alpha(\rv)$ of the nonlinear crystal \cite{opt-modulated}. As $\alpha(\rv)$ inherits the periodicity of the atomic lattice, $\delta\rho$ can be decomposed into a sum over the atomic reciprocal lattice vectors $\Gv$. In particular, $\delta\rho(\rv, t) \approx -i\sum_\Gv \alpha_\Gv (\Ev_\text{i}(\rv, t)\cdot\Gv) \exp(i\Gv\cdot\rv)$, where $\alpha_\Gv$ is the polarizability of the bonds associated with a specific atomic reciprocal lattice vector $\Gv$. More details can be found in the Supplemental Information.

In practice, XO nonlinear processes are driven by strong coherent sources from synchrotrons or FELs \cite{glover, schori, sofer}. As such, the vector potential operator can be decomposed into a sum of a classical pump and a quantum signal as $\Av = \Av_\text{p} + \Av_\text{s}$. In this case, the positive frequency component of the vector potential is $\Av_\text{p} \equiv A_p \hat{\epsilon}_p \exp(i(\kv_p\cdot\rv - \omega_p t))$, a complex-number valued field with amplitude $A_p$, wavevector $\kv_p$, and polarization $\hat{\epsilon}_p$, while the signal and idler fields $\Av_\text{s}$ and $\Ev_\text{i}$ are preserved as quantum operators which encode the quantum state of the emitted photon pair\myfootnote{Specifically, the X-ray signal field is the quantized vector potential operator in vacuum, while the idler field is the quantized electric field operator in the dielectric medium defined by $\varepsilon(\rv)$. Full details are shown in the S.I.}. Keeping only terms which describe the pump-signal interaction results in the XOPDC Hamiltonian
\begin{multline}
    H_{\text{XOPDC}}(t) = -i\frac{q}{m}\sum_\Gv \alpha_\Gv\int_V \text{d}^3r~e^{i\Gv\cdot\rv}(\Ev_\text{i}(\rv, t)\cdot\Gv) \\
    \times(\Av_\text{p}(\rv, t)\cdot\Av_\text{s}(\rv, t)).
    \label{eqn:hint_xopdc}
\end{multline}
Similar forms can be obtained for SFG and DFG, provided that the quantum idler operator is replaced with a coherent complex-valued optical field.

By using Fermi's Golden Rule, we compute the rate $\Gamma$ of the transition from zero signal and idler photons to one of each. We find that for a particular idler mode $i$, the rate of XOPDC per signal frequency and solid angle is
\begin{multline}
    \frac{\text{d}\Gamma_{i,s}}{\text{d}\omega_s \text{d}\Omega_s} = \frac{\omega_i\omega_s^3E_p^2}{16\pi^2c^3} |\hat\epsilon_p\cdot\hat{\epsilon}^*_s|^2 \sum_\Gv |\chi^{(2)}_\Gv|^2\\\left|\int_V d^3r (\hat{G}\cdot\mathbf{F}^*_i(\rv))e^{i(\Gv+\kv_p-\kv_s)\cdot\rv}\right|^2\delta(\omega_p - \omega_s - \omega_i).
    \label{eqn:rate}
\end{multline}
Here, $\chi^{(2)}_\Gv = q\alpha_\Gv |\Gv|/(m\varepsilon_0\omega_p\omega_s)$ is the second-order nonlinearity associated with the nonlinear crystal reciprocal lattice vector $\Gv$, $\mathbf{F}_i(\rv)$ is the idler mode, and $E_p = \omega_p A_p$ is the pump electric field strength\myfootnote{$\mathbf{F}_i(\rv)$ is a normalized eigenfunction of Maxwell's equations in the nanostructure (see S.I. for more details).}. The total detected XOPDC rate is then calculated by summing over the relevant angles, frequencies, and all signal and idler polarizations. Note that the delta function enforces energy conservation between the pump, signal, and idler. In addition, in the limit of a large interaction volume, the phase matching condition for a characteristic idler wavevector $\kv_i$ reduces to $\Gv + \kv_p = \kv_s + \kv_i$.
\begin{figure}
    \centering
    \includegraphics[width=\linewidth]{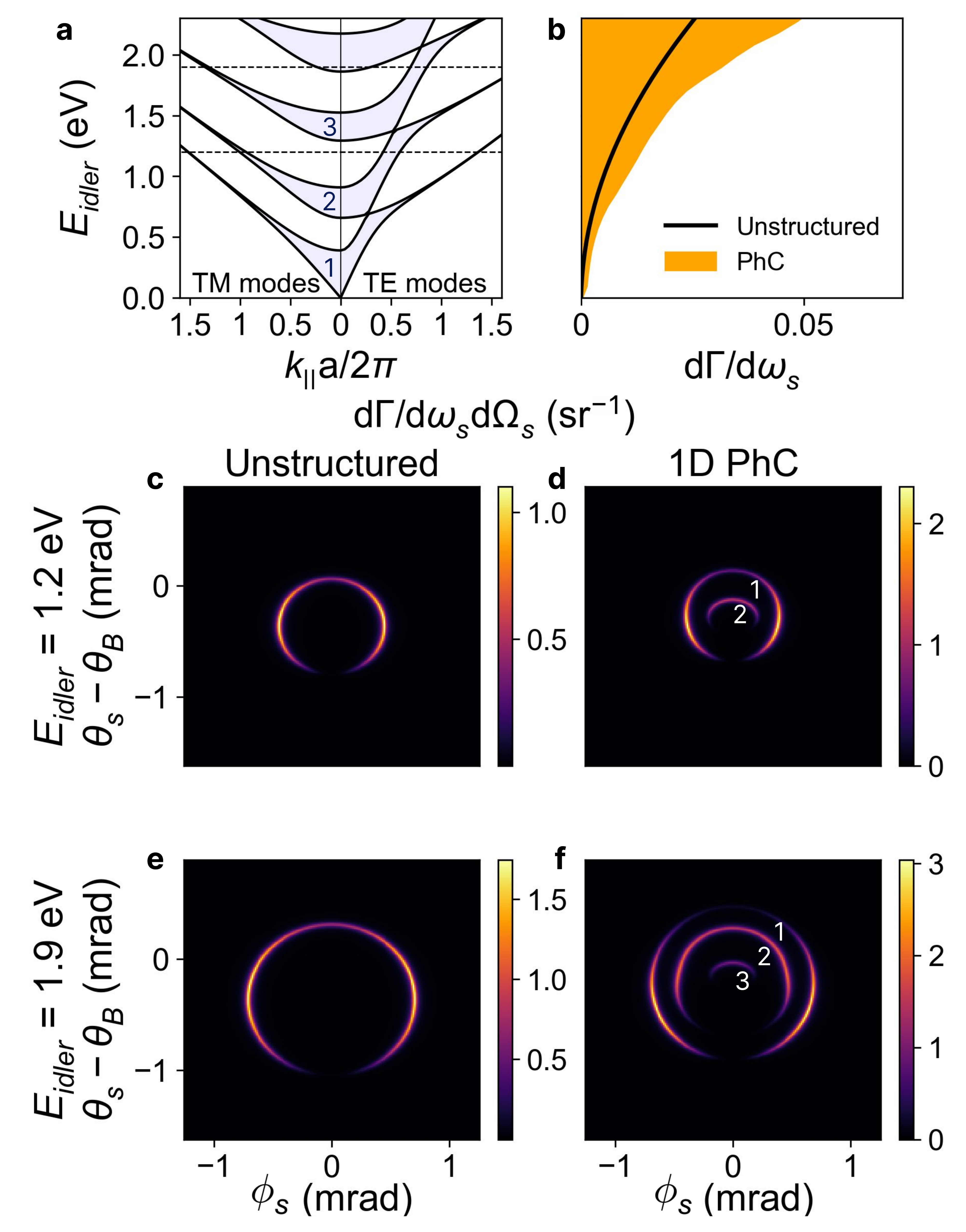}
    \caption{XOPDC in a 1 mm$^3$ gallium arsenide 1D PhC with a pump energy of 6.6 keV. (a) Band structure of the PhC transverse electric/magnetic (TM/TE) modes as a function of the magnitude of the wavevector $k_\parallel$ parallel to the slab surfaces. The allowed regions for each band are outlined in black and filled with gray. The first 3 bands are labeled, and the dashed lines mark idler energies used for panels (c-f). (b) XOPDC rate per unit frequency, normalized by the PhC fill fraction, as a function of the idler photon energy for a 1D PhC and unstructured medium of the same nonlinear material volume and effective index of refraction. (c-f) X-ray angular spectra from unstructured sample and 1D PhC crystal corresponding to idler energies of 0.6 eV and 1.6 eV. At the higher idler energy (f), three emission rings are labeled with numbers that correspond to band numbers shown in (a). Note that plots (c)-(f) are conditioned on the idler energy, and, without idler filtering, the measured X-ray pattern would be averaged over all $\omega_i$.}

    \label{fig:1dphc}
\end{figure}

\begin{figure*}
    \centering
    \includegraphics[width=0.8\textwidth]{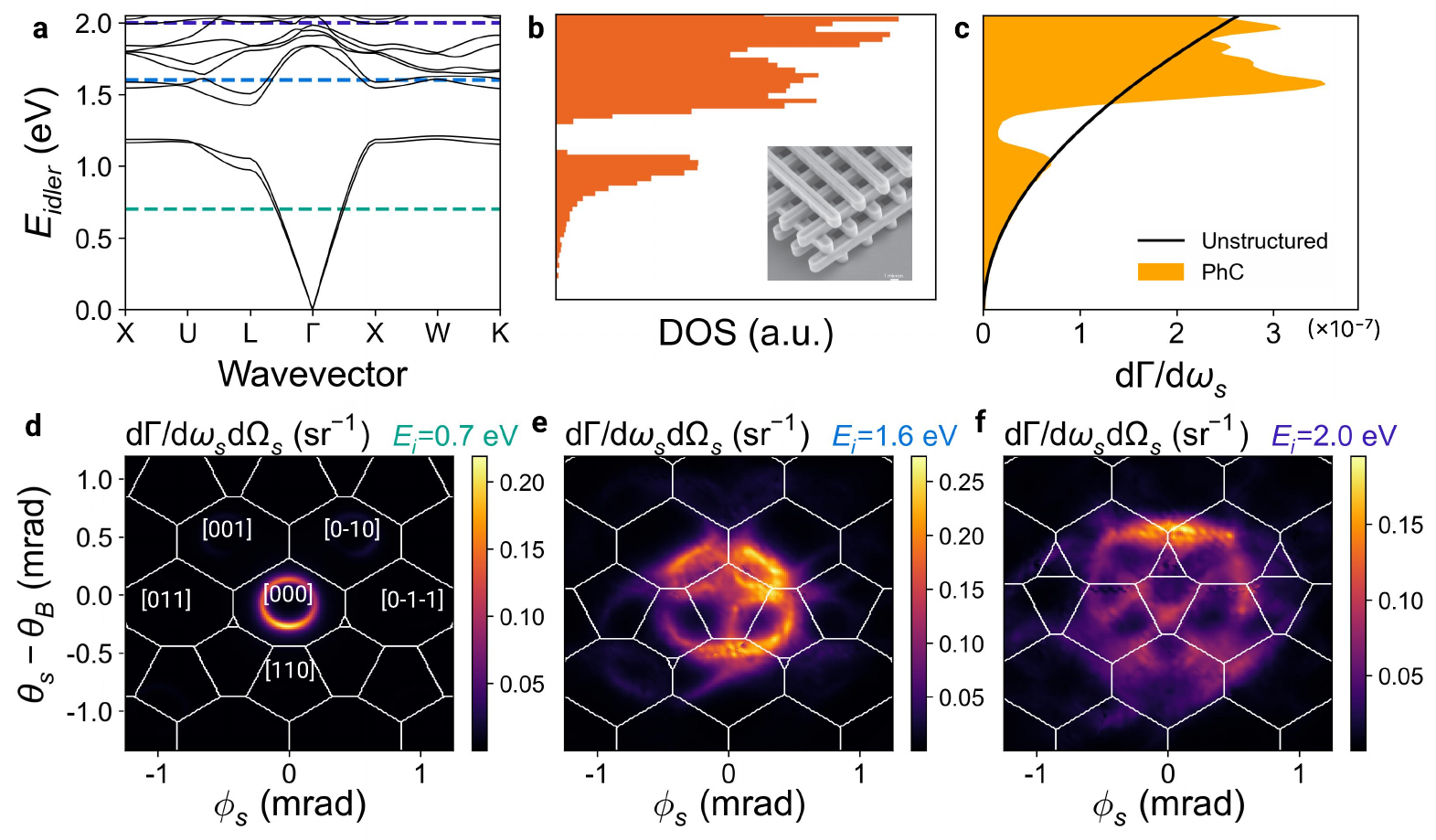}
    \caption{XOPDC in a 3D photonic crystal. (a) Band structure of a gallium arsenide woodpile PhC with a lattice constant of $0.5\mu$m, with symmetry points labelled. Note the complete bandgap from 1.1-1.5 eV. Colored lines indicate the energy levels at which the angular spectra have been plotted. (b) Density of states (DOS) of the PhC. Inset: Electron microscope image of a silicon woodpile crystal from \cite{phcs}. (c) Differential rate of emission of signal/idler pairs in the PhC as a function of the idler photon energy. The black line indicates the emission rate from an unstructured nonlinear crystal with the same volume of nonlinear material and effective index of refraction. (d), (e), (f) X-ray emission spectra in units of photons/s per unit frequency per unit solid angle, conditioned on idler photon energy. The boundaries between X-rays which have diffracted off of different reciprocal lattice vectors of the PhC are denoted by the white lines, with some vectors labeled in (d) in the PhC reciprocal lattice basis. For low idler energy (d), the angular spectrum appears similar to that in an unstructured homogeneous medium (Fig. \ref{fig:1dphc}c and e). At higher idler energies (e, f), the spectrum is symmetric across the reciprocal lattice diffraction boundaries.}
    \label{fig:wdpl}
\end{figure*}

We first illustrate how a 1D PhC shapes the frequency and angular spectrum of the X-rays emitted through XOPDC. A 1D PhC increases the rate of XOPDC at certain frequencies by increasing the DOS available to the idler. This increase in rate is apparent in Fig. \ref{fig:1dphc}b, which shows the rate of signal/idler pair emission per unit frequency in a gallium arsenide (GaAs) 1D PhC consisting of alternating equally thick regions of air and GaAs, summed over all signal and idler polarizations. We consider a PhC with lattice constant $a=0.5~\mu$m, pump photons with an energy of 6.6 keV, and a pump electric field strength of $E_p = 6\times10^{9}$ V/m, which is comparable to that in a state-of-the-art synchrotron. The value of $\chi^{(2)}_\Gv$ for GaAs XOPDC is $4.7\times10^{-14}$ m/V \cite{sofer}. We find that XOPDC in a 1D PhC occurs at a higher rate than in a homogeneous crystal of the same nonlinear material volume and effective index of refraction, leading to an enhancement of 1.9 times. To understand this result, we examine the PhC band structure (Fig. \ref{fig:1dphc}a). With increasing idler energy, more bands become available for emission of an optical photon, thus increasing the optical density of states. For instance, the appearance of the second band around 0.75 eV is accompanied by a rise in the rate of emission, surpassing the rate in an unstructured medium.

The 1D PhC also impacts the angular spread of X-rays emitted through XOPDC. In an unstructured medium, the angular spectrum of X-rays at a particular frequency is approximately dictated by Bragg scattering. While elastic Bragg scattering results in a beam emitted at the Bragg angle $\theta_B = \arcsin\big(\frac{|\Gv|}{2|\kv_p|}\big)$, XOPDC from an unstructured medium results in rings centered around $\theta_B$ with variations due to the dot product of the crystal lattice reciprocal vector and idler polarization from $|\hat{G} \cdot \mathbf{F}_i^*|^2$ in Eq.~\ref{eqn:rate} (Fig.~\ref{fig:1dphc}c,e). Physically, this originates from the system's preference to emit idler photons with wavevectors perpendicular to the reciprocal lattice vector $\mathbf{G}$. For a particular signal-idler frequency pair, the ring radius is set by idler momentum. For idler energies above the first band, such as the ones seen in Fig.~\ref{fig:1dphc}, the X-ray emission from the 1D PhC is concentrated into rings corresponding to the available bands. The origin of this effect is due to different bands having distinct values of the momentum vector parallel to the plane of the PhC ($\kv_{||}$) at the same idler energy. 
 
The properties of emitted X-rays can be further tailored using more complex structures, such as 3D PhCs. These structures afford more control over the DOS at idler frequencies, presenting an opportunity for more precise control of XOPDC rates. To illustrate this, we show results for XOPDC from a GaAs woodpile 3D PhC, which has a complete bandgap (Fig. \ref{fig:wdpl}a). Once again, we sum over all signal and idler polarizations. The rate of XOPDC closely follows the optical DOS (see Figs. \ref{fig:wdpl}b,c). Depending on the DOS, the emission rate for particular idler frequencies can be suppressed and enhanced compared to an unstructured medium. An enhancement of 2.2 is observed at 1.6 eV between the PhC emission and a bulk crystal with the same volume of nonlinear material and effective index of refraction. Almost no X-ray emission should occur for signal-idler pairs in the bandgap. The sharp drop in the DOS which occurs between the band edges and the gap suggests that it may be possible to observe a many-fold change in the X-ray intensity by shifting the X-ray energy detected by only a fraction of an eV. It is important to note that the rate of emission of the X-ray signal does not depend on the index of refraction at the idler wavelength, and increasing the index to increase the optical DOS would not lead to a corresponding rate enhancement at X-ray frequencies (see S.I. for more details).

Due to its three dimensional lattice structure, a 3D PhC can strongly shape the angular spectrum of emitted X-rays. As in a 1D PhC, the X-ray angular spectrum for low energy idlers resembles that of an unstructured sample which produces a ring-shape at the Bragg angle (Fig. \ref{fig:wdpl}d). At higher energies, the shape deviates from the ring. It is well known that a spatially periodic structure, such as a grating, can impart additional integer-spaced momentum onto a photon, leading to diffractive effects. In the context of XOPDC, the PhC reciprocal lattice vectors provide a momentum boost which can shift emission further away from the Bragg angle. 

Mathematically, this phenomenon stems from the fact that the idler wavevector can be written as $\kv_i = \kv_B + \gv$, where $\kv_B$ is in the first Brillouin zone, and $\gv = \ell\mathbf{b}_1 + m\mathbf{b}_2 + n \mathbf{b}_3$ is a PhC reciprocal lattice vector expressed as an integer combination of the reciprocal lattice basis vectors $\mathbf{b}_1, \mathbf{b}_2, \mathbf{b}_3$. For a woodpile PhC, the real-space and reciprocal lattices are face- and body-centered cubic, respectively. Then, the diffraction effect manifests through the phase-matching condition $\kv_p + \Gv = \kv_s + \kv_B + \gv$. As a consequence, every point in angular space $(\theta_s, \phi_s)$ is associated with a single reciprocal lattice vector $\gv$, notated by Miller indices $[\ell mn]$. In the angular spectrum plots, white lines delineate boundaries between regions with different indices. As the idler moves away from the effectively homogenized low-energy region and toward higher energies, the angular emission develops a stronger signature of the PhC band structure (Figs. \ref{fig:wdpl}e-f). In particular, strong emission occurs outside of the Bragg angle region [000]. These distinct angular distributions, which depend heavily on frequency and PhC structure, may provide a way to observe the influence of PhCs on XOPDC.

The experimental realization of these phenomena depends on the interplay between X-ray pump sources, nanophotonic structures which contain XO nonlinear materials, and angle and/or frequency resolved X-ray detection schemes. Fortunately, our calculations indicate that control of XOPDC with PhCs should be possible using schemes similar to those employed to observe XO mixing \cite{glover,schori}. On the pump side, the parameters for incident X-ray intensities are consistent with the specifications of a current X-ray free electron laser \cite{slac-xfel, glover}. On the nanostructure side, optical-IR PhCs have been fabricated from materials which exhibit XO nonlinearity \cite{high-q-ln-phc-nanocav, 3dphcs-ln, gaas-phc-cav, phc-slabs-ln, woodpile-fab}. For instance, a silicon woodpile 3D PhC very similar to the one we considered was fabricated using nanofabrication technology \cite{woodpile-fab}. Existing X-ray detectors can resolve the energy and angular spectral features caused by XOPDC in a PhC. An X-ray monochromator with a bandwidth of 0.25 meV has been fabricated \cite{xray-monochromator}, and state-of-the-art X-ray detectors have resolutions of as low as 25 meV and 55 $\mu$rad \cite{xray-spectrometer-esr}. We estimate that at an energy bin size of 30 meV, collecting X-rays from XOPDC in a gallium arsenide woodpile 3D PhC over the emission region, a few mrad around the Bragg angle, should yield on the order of 10 photons per second, well within the sensitivity of X-ray detectors. Additional developments in either the design of nanostructures or the sensitivity of X-ray instruments would further improve experimental prospects.


We have shown that the X-ray to optical coupling inherent to electronic systems enables the use of optical nanostructures to control X-rays. As an example, we demonstrated how PhCs made of nonlinear crystals can be used to strongly control the frequencies and directions of X-rays emitted through XOPDC, and they can enhance emission rates by 2.2 times over an unstructured nonlinear crystal. As shaping the behavior of optical photons has become routine in nanophotonics, our concept offers one new solution to the challenge of controlling X-rays. Since XOPDC generates entangled X-ray-optical pairs, our work points toward quantum X-ray sources which can be controlled through optical nanostructures. These developments can be enabled by our Hamiltonian framework which describes the quantum optics of XO nonlinearities.
 
Looking forward, the ability to control XO nonlinear processes using optical nanostructures offers opportunities to better probe microscopic scale optical responses of materials and to design new, more finely tunable and monochromatic X-ray sources. For instance, entanglement between X-ray and optical frequencies may enable ghost imaging of lattice and electronic dynamics simultaneously \cite{ghost-imag, two-phot-imag, xray-ghost-imag,valence-probing, nonl-opt-xray-theory}. Furthermore, this coupling may allow for new types of X-ray sources which can be heralded by measuring the emitted optical photon \cite{ent-phot-pairs-spdc}. As our platform shows a path towards creating a source of squeezed, coherent X-ray photons, it may enable an X-ray analog of the optical parametric oscillator, which would be a step towards a much-anticipated more stable and monochromatic source of X-rays. The squeezing in such light sources could also enable spectroscopy of core-shell and nuclear transitions at X-ray frequencies, beyond the standard quantum limit \cite{quant-enh-det, squeezed-light-10db, valence-probing, core-hole-correlation}.  More broadly, we anticipate that our framework will be of use for designing experiments and devices which extend quantum optics into the X-ray regime.

\section*{Acknowledgements and Disclosures}
This material is based upon work supported by the National Science Foundation Graduate Research Fellowship under Grant No. 2139433. This material is based upon work also supported in part by the Defense Advanced Research Projects Agency (DARPA) under Agreement No. HR00112090081, and also in part by the U.S. Army DEVCOM ARL Army Research Office through the MIT Institute for Soldier Nanotechnologies under Cooperative Agreement number W911NF-23-2-0121.  N.R. acknowledges the support of a Junior Fellowship from the Harvard Society of Fellows. We thank Dr. Sharon Shwartz for useful conversations. The authors do not declare any competing interests.

\newpage
\bibliographystyle{apsrev4-1} 
\bibliography{bibliography}
\end{document}